\documentclass[amsfonts,floatfix]{revtex4}
\usepackage{amsmath}
\usepackage{bm}
\usepackage{graphicx}
\usepackage{mathrsfs}
\usepackage{footmisc}
\usepackage{multirow}
\usepackage{graphicx}
\usepackage{booktabs, ctable}

\usepackage[normalem]{ulem}
\usepackage{color}

\renewcommand\sout{\bgroup \color{red} \ULdepth=-.5ex \ULset}

\begin{document}

\title{ Cold and Hot Nuclear Matter Effects on Charmonium Production\\ in p+Pb Collisions at LHC Energy}
\author{Baoyi Chen$^1$, Tiecheng Guo$^{1,2}$, Yunpeng Liu$^1$, Pengfei Zhuang$^2$}
\affiliation{$^1$Physics Department, Tianjin University, Tianjin 300352, China\\
$^2$Physics Department, Tsinghua University, Beijing 100084, China }
\date{\today}

\begin{abstract}
We study cold and hot nuclear matter effects on charmonium production in p+Pb collisions at $\sqrt{s_\text{NN}}=5.02$ TeV in a transport approach. At the forward rapidity, the cold medium effect on all the $c\bar c$ states and the hot medium effect on the excited $c\bar c$ states only can explain well the $J/\psi$ and $\psi'$ yield and transverse momentum distribution measured by the ALICE collaboration, and we predict a significantly larger $\psi'$ $p_\text{T}$ broadening in comparison with $J/\psi$. However, we can not reproduce the $J/\psi$ and $\psi'$ data at the backward rapidity with reasonable cold and hot medium effects.
\end{abstract}
\maketitle

There are two kinds of nuclear matter effects on charmonium production in heavy ion collisions~\cite{review}. One is the hot nuclear matter effect during the evolution of the fireball, and the other is the cold nuclear matter effect before the formation of the fireball. The former includes color screening~\cite{satz} and regeneration~\cite{thews,andronic,grandchamp,yan} which work in an opposite way and lead respectively to charmonium suppression and enhancement. The later contains mainly the shadowing effect~\cite{eks,vogt,fs,eps}, Cronin effect~\cite{cronin,gavin} and nuclear absorption~\cite{gerschel}. To be sure that the charmonium production is a sensitive probe of the hot Quark-Gluon Plasma (QGP)~\cite{review}, one needs to clearly understand the cold medium in nucleus-nucleus (A+B) collisions. Proton-nucleus (p+A) collisions where the hot medium is expected to be small and the effect is weak are widely considered as a good laboratory to measure the cold medium.

Recently, the ALICE collaboration measured the nuclear modification factor $R_\text{pA}$ and averaged transverse momentum square $\langle p_\text{T}^2\rangle_\text{pA}$ for $J/\psi$ and $\psi'$ in p+Pb collisions at $\sqrt{s_{NN}}=5.02$ TeV~\cite{alice1,alice2,alice3,alice4}. The charmonium production in small systems is also widely investigated in energy loss model~\cite{arleo}, transport model~\cite{du} and comover model~\cite{ferreiro}. In this paper, we focus on the charmonium evolution in p+A collisions in the frame of a transport approach~\cite{zhu,yan}. By comparing with the data, we hope to understand the cold and hot nuclear matter effects on the ground and excited charmonium states.

There are different charmonium production mechanisms, initial production via hard processes~\cite{gerschel2}, recombination of charm quarks inside QGP~\cite{thews,andronic,grandchamp,yan}, and decay from excited states and B hadrons~\cite{cdf,cms,chen}. Since the charm quark number created in p+A collisions is small, the recombination can be reasonably ignored. At LHC energy,
the collision time is much shorter than the charmonium formation time and the QGP formation time. Therefore, one can safely neglect
the nuclear absorption and the difference in the shadowing and Cronin effects between the ground and excited charmonium states.

Generally, the initially produced charmonium distribution $f_\Psi$ for $\Psi=J/\psi,\ \psi',\ \chi_c$ in transverse plane in an A+B collision at fixed impact parameter ${\bf b}$ can be obtained through a superposition of effective nucleon-nucleon (p+p) collisions~\cite{zhou},
\begin{equation}
\label{initial}
f_\Psi({\bf p},{\bf x}|{\bf b}) = (2\pi)^3\delta(z)\int dz_\text{A}dz_\text{B}\rho_\text{A}({\bf x}_T,z_\text{A})\rho_\text{B}({\bf x}_T-{\bf b},z_\text{B})
{\cal R}_{\text g}(x_1,\mu_{\text F},{\bf x}_{\text T}){\cal R}_{\text g}(x_2,\mu_{\text F},{\bf x}_{\text T}-{\bf b}){d\overline \sigma_\Psi^\text {pp}\over d^3{\bf p}},
\end{equation}
where ${\bf x}_\text{T}$ is the charmonium transverse coordinate, and $\rho_\text{A,B}$ are the nucleon distribution functions in nuclei A and B. We have here taken the gluon fusion $g+g\rightarrow g+\Psi$ as the dominant hard process, and $z_\text{A,B}$ are the longitudinal coordinates of the two nucleons in nuclei A and B where the two gluons come from. The shadowing effect is embedded in the inhomogeneous modification factor ${\cal R}_\text{g}$~\cite{vogt} for the initial gluons with longitudinal momentum scale $x=E_\text{T}/\sqrt{s_{\text{NN}}}\ e^y$ and factorization factor $\mu_\text{F}=E_\text{T}$~\cite{eks}, where $E_\text{T}=\sqrt{m_\Psi^2+{\bf p}_\text{T}^2}$ and $y=1/2\ln((E+p_\text{z})/(E-p_\text{z}))$ are the charmonium transverse energy and longitudinal momentum rapidity. We employ the EPS09 NLO~\cite{eps,eps2} to parameterize the homogeneous shadowing effect in our numerical calculation.

The effective cross section $\overline \sigma_\Psi^\text {pp}$ includes the Cronin effect on charmonium production in nuclear collisions. Before two
gluons fuse into a charmonium, they acquire additional transverse
momentum via multi-scattering with the around nucleons,
and this extra momentum is inherited by the produced
charmonium. Inspired from a random-walk picture, we
take a Gaussian smearing~\cite{zhao,chen2} for the modified transverse
momentum distribution
\begin{equation}
\label{cronin}
\overline f^{\text{pp}}_\Psi({\bf p},{\bf x}_{\text T},z_{\text A},z_{\text B})={1\over \pi a_{\text{gN}} l} \int
d^2{\bf q}_\text{T} e^{-{\bf q}_\text{T}^2\over a_{\text{gN}} l}f^{\text{pp}}_\Psi(|{\bf
p}_\text{T}-{\bf q}_\text{T}|,p_\text{z}),
\end{equation}
where $l({\bf x}_\text{T},z_\text{A},z_\text{B})$ is the averaged path length of the two gluons coming from $z_\text{A}$ and $z_\text{B}$ in
nuclei before their fusion into a charmonium at ${\bf x}_\text{T}$, $a_{\text{gN}}$ is the averaged charmonium transverse
momentum square obtained from the gluon scattering with a unit of length of nucleons, and $f^{\text{pp}}_\Psi({\bf p})$ is the
momentum distribution for a free p+p collision. The length
$l$ is calculated from the nuclear geometry, and the value of the Cronin parameter
is extracted as $a_{\text{gN}}=0.15\ \mathrm{GeV^2/fm}$ by fitting the data in $\sqrt{s_{\text{NN}}}=2.76$ TeV Pb-Pb collisions~\cite{chen2}.
We take the same value  at
$\sqrt{s_{\text{NN}}}=5.02$ TeV.

There is no experimental data for the distribution $f_\Psi^{\text{pp}}$ in free p+p collisions at $\sqrt{s_{\text{NN}}}=5.02$ TeV. We parameterize the distribution with experimental data at $\sqrt{s_{\text{NN}}}=2.76$ TeV~\cite{alice5}
and $7$ TeV~\cite{alice6} as
\begin{equation}
{d^2\sigma_\Psi^{\text{pp}}\over dyp_\text{T}dp_\text{T}} = {(n-1)\over \pi (n-2)\langle p_\text{T}^2\rangle_\Psi^{\text{pp}}}
\left(1+{p_\text{T}^2\over (n-2)\langle p_\text{T}^2\rangle_\Psi^{\text{pp}}}\right)^{-n} {d\sigma_\Psi^{\text{pp}}\over dy}
\end{equation}
with $n=3.2$, $d\sigma_\Psi^{\text{pp}}/dy=5.01 e^{-0.06418 y^2} \mu$b, the
averaged transverse momentum square~\cite{alice5,alice6,phenix} $\langle p_\text{T}^2\rangle _\Psi^{\text{pp}} (y)= 12.5 \left(1-(y/y_{\text{max}})^2\right)$ (GeV/c)$^2$, and the maximal charmonium rapidity $y_{\text{max}}=\ln(\sqrt{s_{\text{NN}}}/m_\Psi)$ in p+p collisions.

With the phase-space distribution (\ref{initial}) as the initial condition at time $\tau_0$ where the fireball is formed, the evolution of the initially produced charmonia in the hot medium can be described by the transport equation~\cite{liu2}
\begin{equation}
\left[\cosh(y-\eta){\frac{\partial}{\partial\tau}}+{\frac{\sinh(y-\eta)}{\tau}}{\frac{\partial}{\partial
\eta}}+{\bf v}_\text{T}\cdot\nabla_\text{T}\right]f_\Psi=-\alpha_\Psi f_\Psi
\label{tran}
\end{equation}
where ${\bf v}_\text{T}={\bf p}_\text{T}/E_\text{T}$ and $\eta=1/2\ln((t+z)/(t-z))$ are respectively the charmonium transverse velocity and longitudinal coordinate rapidity. The third term on the left hand side represents the leakage effect on charmonia in the transverse plane. The charmonia with large velocity
can escape from the hot medium by free streaming.
This effect will increase the charmonium transverse momentum
in nuclear collisions~\cite{zhu}. The charmonium suppression in the hot medium is reflected in the loss term $\alpha_\Psi$~\cite{yan},
\begin{equation}
\label{alpha}
\alpha_\Psi({\bf p},{\bf x},t,T) ={1\over 2E_\text{T}} \int {d^3{\bf k}\over
{(2\pi)^3 2E_\text{g}}}F_{\text{g}\Psi}({\bf p},{\bf k})\sigma_{\text{g}\Psi}({\bf p},{\bf k},T)f_\text{g}({\bf k},T),
\end{equation}
where we have considered the gluon dissociation $g+\Psi\to c+\bar c$ as the dominant dissociation process in the hot QGP, and $E_\text{g}$ and $\bf k$ are the gluon energy and momentum. The cross section $\sigma_{\text{g}\Psi}$ in vacuum can be derived through the operator production expansion with perturbative Coulomb wave function~\cite{peskin,bhanot}, and its temperature dependence can be obtained by taking into account the geometrical relation between the cross section and the average size of the charmonium state, $\sigma_{\text{g}\Psi}(T) =\sigma_{\text{g}\Psi}(0)\langle r_\Psi^2\rangle(T)/\langle r_\Psi^2\rangle(0)$. The averaged radius squared $\langle r_\Psi^2\rangle$ can be calculated from the potential
model~\cite{satz2}, its divergence self-consistently defines a Mott dissociation temperature $T_\text{d}^\Psi$ which indicates the melting of the bound state due to color screening. In this sense, the cross section in medium can also be approximately written as $\sigma_{\text{g}\Psi}(T)=\sigma_{\text{g}\Psi}(0)/\Theta(T_\text{d}^\Psi-T)$.

The thermal gluon distribution $f_\text{g}=16/(e^{k_\mu u^\mu/T}-1)$ in (\ref{alpha}) is controlled by the local temperature $T({\bf x},t)$ and fluid velocity $u_\mu({\bf x},t)$ which are determined by the hydrodynamics of the hot medium,
\begin{equation}
\partial_\mu T^{\mu\nu} = 0
\label{hydro}
\end{equation}
with the energy-momentum tensor $T^{\mu\nu}$. In order to close the hydrodynamic equations, one needs the equation of state of the medium. We follow Ref.~\cite{sollfrank}
where the deconfined phase at high temperature is an ideal gas of gluons and massless u and d quarks plus 150 MeV mass s quarks, and the hadron phase at low temperature is an
ideal gas of all known hadrons and resonances with mass up to 2 GeV~\cite{hagiwara}. There is a first-order phase transition between
these two phases. In the mixed phase, the Maxwell construction is used. The mean field repulsion parameter and the bag
constant are chosen as $K=450$ $\mathrm{MeV\ fm^3}$~\cite{sollfrank} and $B^{1/4}=236$ MeV to obtain the critical temperature $T_\text{c}=165$ MeV at vanishing baryon density.

\begin{figure}[!hbt]
\centering
\includegraphics[width=0.35\textwidth]{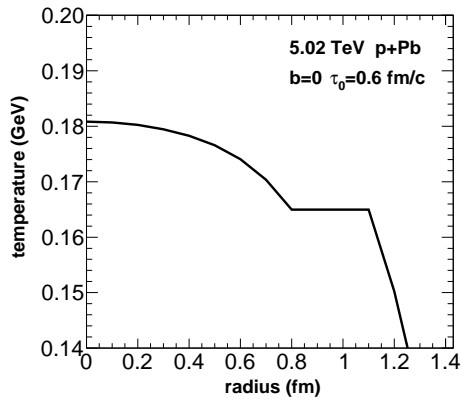}
\caption{The initial temperature distribution as a function of the fireball radius in central p+Pb collisions at colliding energy $\sqrt{s_{\text{NN}}}=5.02$ TeV. }
\hspace{-0.1mm}
\label{fig1}
\end{figure}

We choose the initial time of the hot medium to be $\tau_0=0.6$ fm~\cite{liu3}, and determine the spatial dependence of the initial energy density by the Glauber model. By taking the maximum temperature at the center of the fireball as a free parameter and fitting its value to be $180$ MeV, the profile of the initial temperature in central p+Pb collisions at $\sqrt{s_{\text{NN}}}=5.02$ TeV is shown in Fig.\ref{fig1}. The initial hot medium is in the QGP phase in the region of fireball radius $0<r<0.8$ fm and enters the mixed phase at 0.8 fm $<r<$ 1.1 fm where the QGP fraction changes from 1 to 0. The temperature jumps down rapidly afterwards.

The inclusive charmonia measured by the ALICE Collaboration include the prompt part and the contribution from the B decay. The former consists of direct production and feed-down from the excited states, $30\%$ from $\chi_c$ and $10\%$ from $\psi'$, and the latter comes from the decay of bottomed hadrons. Since the ALICE experiment does not separate the two parts from each other, we have to take into count the B decay contribution in our model to compare our theoretical calculation with the ALICE data. From the inclusive $J/\psi$ data in p+p collisions, the transverse momentum dependence of the B decay fraction can be well parameterized as $f_{\text{B}\rightarrow \Psi}=f_0+k p_\text{T}$ with $f_0=0.04$ and $k=0.023\ c/\textrm{GeV}$ for $J/\psi$~\cite{cms2} and $f_0=0.114$ and $k=0.0217\ c/\textrm{GeV}$ for $\psi'$~\cite{cdf}. Note that the linear
parametrization is rapidity independent in the region we considered.

\begin{figure}[!hbt]
\centering
\includegraphics[width=0.35\textwidth]{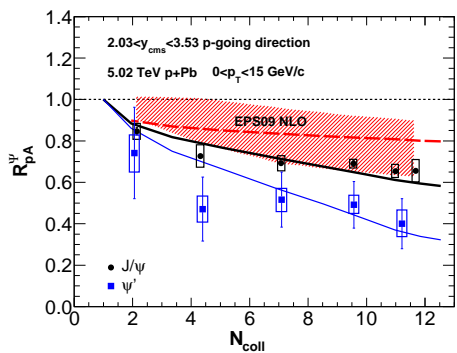}
\caption{(Color online) The charmonium nuclear modification factor $R_{\text{pA}}^\Psi$ as a function of the number of binary collisions $N_{\text{coll}}$ in p+Pb collisions at $\sqrt{s_{\text{NN}}}=5.02$ TeV at forward rapidity. The dashed line shows the result with only cold nuclear matter effect parameterized by EPS09 NLO~\cite{eps,eps2}, the uncertainty in shadowing is represented by the band~\cite{alice4}. The thick and thin solid lines are respectively the full calculations for inclusive $J/\psi$ and $\psi'$, and the data are from the ALICE Collaboration~\cite{alice4,alice7}. }
\hspace{-0.1mm}
\label{fig2}
\end{figure}

Including the contributions from the direct production controlled by the transport equation (\ref{tran}) and the decay from excited states and B hadrons, we now calculate the charmonium distributions in the final state and compare them with the ALICE data. We start with the charmonium nuclear modification factor $R_{\text{pA}}^\Psi=N_{\text{pA}}^\Psi/(N_{\text{coll}}N_{\text{pp}}^\Psi)$, where $N_{\text{coll}}$ is the number of binary collisions which can be calculated from the nuclear geometry at fixed centrality, and $N_{\text{pp}}^\Psi$ and $N_{\text{pA}}^\Psi$ are respectively the numbers of charmonia measured in p+p and p+A collisions. Our calculated $R_{\text{pA}}$ as a function of $N_{\text{coll}}$ for $J/\psi$ and $\psi'$ and the comparison with the inclusive ALICE data for p+Pb collisions at $\sqrt{s_{\text{NN}}}=5.02$ TeV are shown in Fig.\ref{fig2} at forward rapidity.

As is well known, the shadowing effect on charmonium production in nuclear collisions depends strongly on the charmonium rapidity $y$. From the relation between the parton momentum fraction and the charmonium rapidity $x\sim e^y$, there is a strong shadowing at forward rapidity and antishadowing at backward rapidity in nuclear collisions at LHC energy. This provides a chance to see clearly the cold nuclear matter effect on charmonium distributions in different rapidity bins. Taking only the shadowing effect parameterized by EPS09 NLO~\cite{eps,eps2} leads already to a sizeable charmonium suppression, see the dashed line in Fig.\ref{fig2}. The band is the result with uncertainty in the shadowing~\cite{alice4}. Note that the cold medium effect is almost the same for the ground and excited $c\bar c$ states. While the maximum shadowing effect (the lower limit of the band) can reasonably explain the $J/\psi$ data, it is not enough to reproduce the $\psi'$ data. The hot medium effect can be estimated from the charmonium dissociation temperature $T_\text{d}^\Psi$. Using the potential model~\cite{satz2} at finite temperature, $T_\text{d}^{J/\psi}\simeq 1.5T_\text{c}$ is much higher than the fireball temperature in p+Pb collisions shown in Fig.\ref{fig1} and the directly produced $J/\psi$s are almost not affected by the hot medium. However, the central temperature of the fireball is higher than $T_\text{d}^{\psi'}\simeq T_\text{d}^{\chi_c}\simeq T_\text{c}$, and $\psi'$s are strongly suppressed by the hot medium. Let us focus on the $J/\psi$ suppression in very central collisions where the hot medium effect becomes the strongest. Assuming that all the direct $J/\psi$s survive the fireball and the suppression is due to the contribution of the excited states. Since the cold and hot nuclear matter effects happen in different time regions, they independently affect the charmonium production. Introducing the cold and hot medium induced modification factors $R_\text{pA}^\text{cold}$ and $R_\text{pA}^\text{hot}$ and considering the fact that $40\%$ of the finally measured $J/\psi$s come from the decay of the excited states $\psi'$ and $\chi_c$, $R_\text{pA}^\Psi$ can be reexpressed as
\begin{eqnarray}
R_{\text{pA}}^{\psi'} &=& R_\text{pA}^\text{cold}R_\text{pA}^\text{hot},\nonumber\\
R_{\text{pA}}^{J/\psi} &=& 0.6R_\text{pA}^\text{cold}+0.4R_\text{pA}^{\psi'}.
 \label{estimation}
\end{eqnarray}
Taking $R_\text{pA}^\text{cold}=0.8$ from the EPS09 NLO parametrization shown in Fig.\ref{fig2} and $R_\text{pA}^\text{hot}=0.5$, we have $R_\text{pA}^{\psi'}=0.4$ and $R_\text{pA}^{J/\psi}=0.64$ which are consistent with the ALICE data and the transport model calculation, see Fig.\ref{fig2}.

The $R_\text{pA}(N_\text{coll})$ as a function of centrality is a global quantity, the momentum integration smears the fireball structure and the cold and hot nuclear matter effects at different energies. To have a deep insight into what is happening to the quarkonium motion in the hot medium, one needs to concentrate on the quarkonium momentum distribution. Different from the longitudinal motion which inherits the initial colliding kinematics via momentum conservation, the transverse motion in heavy ion collisions is developed during the dynamical evolution of the system. The microscopically high particle density and multiple scatterings play an essential role in the development of the finally observed transverse momentum distribution. The distribution is therefore sensitive to the medium properties, such as the equation of state. For quarkonia, we expect that their transverse momentum distribution can help us to probe the detailed structure of the fireball and differentiate between the production and suppression mechanisms.

\begin{figure}[!hbt]
\centering
\includegraphics[width=0.40\textwidth]{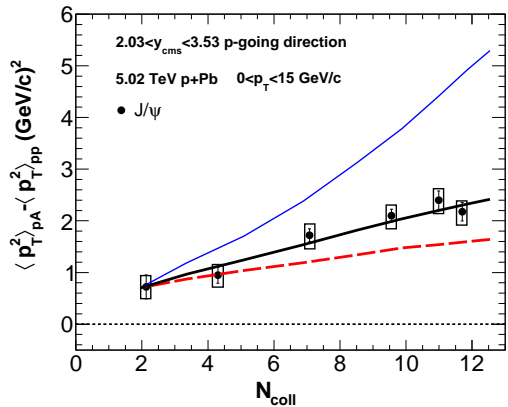}
\caption{(Color online) The nuclear effect induced change in charmonium averaged transverse momentum square $\langle p_\text{T}^2\rangle_{\text{pA}}-\langle p_\text{T}^2\rangle_{\text{pp}}$ as a function of the number of binary collisions $N_{\text{coll}}$ in p+Pb collisions at $\sqrt{s_{\text{NN}}}=5.02$ TeV at forward rapidity. The dashed line shows the result with only cold nuclear matter effect parameterized by EPS09 NLO~\cite{eps,eps2}, and the thick and thin solid lines are respectively the full calculations for inclusive $J/\psi$ and $\psi'$. The data for $J/\psi$ are from the ALICE Collaboration~\cite{alice4}. }
\hspace{-0.1mm}
\label{fig3}
\end{figure}

We now turn to the calculation of the averaged transverse momentum square where the Cronin effect plays an important role. The difference $\langle p_\text{T}^2\rangle_{\text{pA}}-\langle p_\text{T}^2\rangle_{\text{pp}}$ between the p+Pb and p+p collisions is shown in Fig.~\ref{fig3} as a function of $N_{\text{coll}}$. Since there are no data on charmonium transverse momentum distribution in p+p collisions at $\sqrt{s_{\text{NN}}}=5.02$ TeV~\cite{alice4}, we fix the difference at $N_{\text{coll}}=2$ by fitting the data. In the most central collisions, the averaged length $l$ is roughly the radius of Pb, that is 6.6 fm, which leads to an increase of $a_{\text{gN}}l=0.15\times 6.6$ GeV$^2\simeq 1$ GeV$^2$. The shadowing effect which is stronger at low $p_\text{T}$~\cite{alice3} leads to a further $p_\text{T}$ broadening. For the excited state $\psi'$, see the thin solid line in Fig.\ref{fig3}, the hot nuclear matter effect changes its momentum distribution significantly. In principle, low momentum particles are strongly affected by the medium, and fast moving particles are governed by the dynamics in vacuum. This is reflected in both the charmonium suppression rate $\alpha$ and the leakage effect. With the gluon dissociation cross section in the frame of OPE~\cite{peskin,bhanot}, the suppression rate at low $p_\text{T}$ is significantly stronger than that at high $p_\text{T}$~\cite{liu}. Due to the leakage effect, the fast moving $\psi'$s can even escape from the hot medium region before its formation. Since the temperature, size and life time of the fireball increase with centrality, the hot medium induced $p_\text{T}$ broadening is especially important in central collisions. This strong enhancement of averaged transverse momentum square for $\psi'$ can be considered as a signal of the hot medium effect in p+Pb collisions and it needs to be experimentally confirmed.

\begin{figure}[!hbt]
\centering
\includegraphics[width=0.40\textwidth]{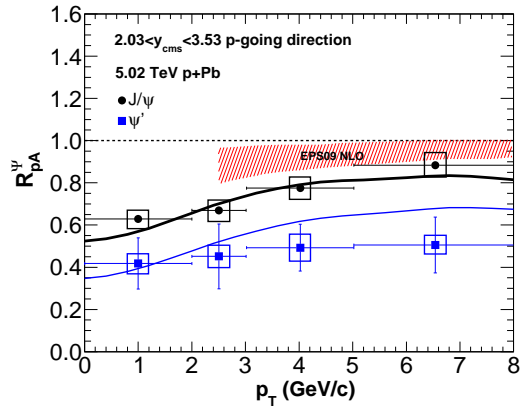}
\caption{(Color online) The charmonium nuclear modification factor $R_{\text{pA}}^\Psi$ as a function of transverse momentum $p_\text{T}$ in minimum bias (impact parameter b=5 fm) p+Pb collisions at $\sqrt{s_{\text{NN}}}=5.02$ TeV at forward rapidity. The band is the result with only cold nuclear matter effect parameterized by EPS09 NLO~\cite{eps,eps2}, the thick and thin solid lines are respectively the full calculations for inclusive $J/\psi$ and $\psi'$, and the data are from the ALICE Collaboration~\cite{alice2}. }
\hspace{-0.1mm}
\label{fig4}
\end{figure}

The differential nuclear modification factor $R_{\text{pA}}(p_\text{T})$ as a function of transverse momentum is shown in Fig.\ref{fig4} for inclusive $J/\psi$ and $\psi'$ in minimum bias (impact parameter b=5 fm) p+Pb collisions at $\sqrt{s_\text{NN}}=5.02$ TeV at forward rapidity.  The band is the result including only cold nuclear matter effect parameterized by EPS09 NLO~\cite{eps,eps2} with uncertainty in shadowing effect, it clearly deviates from the data, especially the data of $\psi'$. The full calculation with both cold and hot nuclear matter effects reproduce well the data. The increasing $R_{\text{pA}}$ with $p_\text{T}$ indicates the strong suppression at low $p_\text{T}$ and weak suppression at high $p_\text{T}$, and the almost $p_\text{T}$ independent difference between the two curves shows that, there is no suppression for direct $J/\psi$ and only the excited states are affected by the hot medium. One can also use Eq.(\ref{estimation}) to estimate the result here. For example, in the low $p_\text{T}$ region with $p_\text{T}>4$ GeV$/c$, we take from Fig.\ref{fig4} the background factor $R_\text{pA}^\text{cold} = 0.95$ and the suppression factor $R_\text{pA}^{\psi'}=0.5$ for $\psi'$, Eq.(\ref{estimation}) leads to the nuclear modification factor $R_\text{pA}^{J/\psi}=0.77$.

\begin{figure}[!hbt]
\centering
\includegraphics[width=0.40\textwidth]{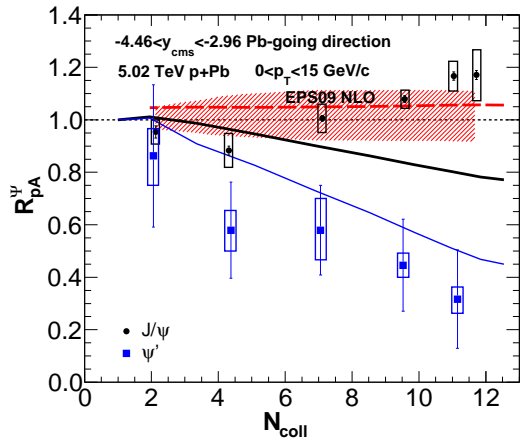}
\caption{(Color online) The charmonium nuclear modification factor $R_{\text{pA}}^\Psi$ as a function of the number of binary collisions $N_{\text{coll}}$ in p+Pb collisions at $\sqrt{s_{\text{NN}}}=5.02$ TeV at backward rapidity. The dashed line shows the result with only cold nuclear matter effect parameterized by EPS09 NLO~\cite{eps,eps2}, the uncertainty in antishadowing is represented by the band. The thick and thin solid lines are respectively the full calculations for inclusive $J/\psi$ and $\psi'$, and the data are from the ALICE Collaboration~\cite{alice4,alice7}. }
\hspace{-0.1mm}
\label{fig5}
\end{figure}

We now extend our calculation to the backward rapidity. The momentum integrated nuclear modification factor $R_\text{pA}$ as a function of $N_\text{coll}$ is shown in Fig.\ref{fig5}. The most big difference between the two rapidity regions is the shadowing at forward rapidity and antishadowing at backward rapidity. This can be seen clearly in Figs.\ref{fig2} and \ref{fig5} where the dashed line is below the unit at forward rapidity but above the unit at backward rapidity. Taking the same hot medium effect as at the forward rapidity, while the calculated modification factor $R_\text{pA}^{\psi'}$ for $\psi'$ at the backward rapidity roughly agrees with the data, the calculation for $J/\psi$ is however in opposition to the data: $J/\psi$ is enhanced in data but suppressed in the calculation. Is it possible to reproduce the $J/\psi$ and $\psi'$ data with reasonable cold and hot medium effects? From Fig.\ref{fig5} and Eq.(\ref{estimation}), to fit $R_\text{pA}^{\psi'}=0.3$ and $R_\text{pA}^{J/\psi}=1.1$ in central collisions, one has to have $R_\text{pA}^\text{cold}=1.63$ and $R_\text{pA}^\text{hot}=0.18$. Taking gluon fusion as the initial $c\bar c$ production process, the gluon modification factor is $R_\text{g} = R_\text{pA}^\text{cold}=1.63$ which is far beyond the upper limit of the normally parameterized antishadowing~\cite{eks,vogt2}. The hot medium factor is also extremely strong and far beyond the expectation for p+A collisions. Note that, even if we can explain the yield of $J/\psi$ and $\psi'$, the regeneration as a consequence of the strong hot medium effect will definitely lead to a decreasing averaged transverse momentum square $\langle p_\text{T}^2\rangle_{pA}-\langle p_\text{T}^2\rangle_{pp} < 0$~\cite{zhou,zhou2}, while the data from the ALICE collaboration increase with centrality $\langle p_\text{T}^2\rangle_{pA} - \langle p_\text{T}^2\rangle_{pp} > 0$~\cite{alice4}.

In our calculations, we did not consider the hot medium effect at hadron level which may play a role in p+A collisions. However, there is still $J/\psi$ suppression $R_\text{pA}^{J/\psi}=0.95$ in central collisions in the calculation with EPS09 LO for cold medium effect ($R_\text{pA}^\text{cold}=1.2$) and comover interaction for hot medium effect~\cite{ferreiro}. When we take EPS09 NLO ($R_\text{pA}^\text{cold}=1.05$) as in our calculation, the obtained $R_\text{pA}^{J/\psi}=0.8$ becomes almost the same as the result shown in Fig.\ref{fig5}.

In summary, we investigated charmonium production in p+Pb collisions at colliding energy $\sqrt{s_\text{NN}}=5.02$ TeV. The cold nuclear matter effect on all $c\bar c$ states and hot nuclear matter effect on the excited $c\bar c$ states only can explain well the charmonium yield and transverse momentum distribution at forward rapidity. The strong $\psi'$ suppression supports the fireball formation in the collisions. In comparison with $J/\psi$, the hot medium effect leads to a significant $p_\text{T}$ broadening for $\psi'$. However, we can not simultaneously reproduce the charmonium $R_\text{pA}$ and $\langle p_\text{T}^2\rangle_\text{pA}$ at backward rapidity with reasonable cold and hot medium effects. There seems to be something new at the backward rapidity.

\appendix {\bf Acknowledgment}:
The work is supported by the NSFC and MOST grant Nos. 11335005, 11575093, 11547043, 2013CB922000, 2014CB845400 and Tsinghua University Initiative Scientific Research Program.

\end{document}